\documentclass[a4paper,fleqn,usenatbib]{mnras}

\usepackage{newtxtext,newtxmath}

\usepackage[T1]{fontenc}
\usepackage{ae,aecompl}

\usepackage[final]{graphicx}	
\usepackage{amsmath}	
\usepackage{amssymb}	
\usepackage{float}
\usepackage{caption}
\usepackage{subfig}

\title[The formation of high-mass binary star systems]{The formation of high-mass binary star systems}

\author[K. Lund \& I. A. Bonnell]{
Kristin Lund,$^{1}$\thanks{E-mail: kblg@st-andrews.ac.uk}
Ian A. Bonnell$^{1}$
\\
$^{1}$SUPA, School of Physics and Astronomy, University of St Andrews, North Haugh, St Andrews KY16 9SS, UK
}

\date{Accepted XXX. Received YYY; in original form ZZZ}

\pubyear{2017}

\begin{document}
\label{firstpage}
\pagerange{\pageref{firstpage}--\pageref{lastpage}}
\maketitle

\begin{abstract}
We develop a semi-analytic model to investigate how accretion onto wide low-mass binary stars can result in a close high-mass binary system. The key ingredient is to allow mass accretion while limiting the gain in angular momentum. We envision this process as being regulated by an external magnetic field during infall. Molecular clouds are made to collapse spherically with material either accreting onto the stars or settling in a disk. Our aim is to determine what initial conditions are needed for the resulting binary to be both massive and close. Whether material accretes, and what happens to the binary separation as a result, depends on the relative size of its specific angular momentum, compared to the specific angular momentum of the binary. When we add a magnetic field we are introducing a torque to the system which is capable of stripping the molecular cloud of some of its angular momentum, and consequently easing the formation of high-mass binaries. Our results suggest that clouds in excess of 1000 M$_\odot$ and radii of 0.5 pc or larger, can easily form binary systems with masses in excess of 25 M$_\odot$ and separations of order 10 R$_\odot$ with magnetic fields of order 100 $\mu$G (mass-to-flux ratios of order 5). 

\end{abstract}




\section{Introduction}

The formation of high-mass stars is a challenging theoretical problem due to the combination of their mass being ten to one hundred times that of a typical star, and their propensity to be found in the crowded centres of stellar clusters \citep{Zinnecker2007}. This implies that either their pre-fragmentation natal conditions are very different from what we expect in star forming regions, or that the bulk of their masses needed to be accumulated in a post-fragmentation accretion phase \citep{Bonnell1998,McKee2003,Bonnell2004}. The fact that many of these stars are also in binary systems with a close, high-mass companion \citep{Mason1998,Zinnecker2001,Sana2014} highlights the difficulty in forming these stars in such crowded systems. Under reasonable physical conditions, the pre-collapse fragments cannot fit into the system.

High-mass stars are a rare, but important contributor to the energetics of the interstellar medium. Their luminous, kinetic and chemical input into the interstellar medium forms a significant part of how galaxies evolve. Models of how high-mass stars form involve either a turbulent core scenario where turbulence is envisioned to support a dense core for many dynamical times and somehow stop it from fragmenting \citep{McKee2003,Dobbs2005,Krumholz2010}, or a competitive accretion model whereby accretion due to the large-scale potential of a stellar cluster allows the stars at the bottom of the potential to grow to become high-mass stars \citep{Bonnell1997,Bonnell2001,Bonnell2004}. 

The fact that many of these high-mass stars are found in very close ($\le 1 au$) binary systems with another high-mass star has proved a complication for all models of high-mass star formation. These high-mass close binaries are of even more interest due to the recent detection of gravity waves from the merger of 30 solar mass binary black holes \citep{Abbott2017}. \cite{Krumholz2010} have suggested a disk fragmentation followed by a common-envelope type event, but it is unclear if such a scenario is feasible to form such close systems. \cite{Bonnell2005} suggested accretion of low-angular momentum material but this would also be limited by the significant non-zero angular momentum in the infalling gas.

In this paper, we explore the possibility that magnetic braking in the infalling gas can act to remove sufficient angular momentum to form a very close high-mass binary. Magnetic braking in pre-cloud collapse \citep{Mouschovias1979} and stellar winds \citep{Mestel1984,Ud-Doula2009} has been evoked in other contexts.
Of particular relevance here is that the gas that accretes onto a forming high-mass star can come from larger distances in the cluster \citep{Bonnell2004} and have lower gas densities \citep{Smith2009} such that there is a greater likelihood of the gas being tied to the magnetic field. Magnetic fields are observed in molecular clouds with strengths that indicate they can be relevant to the accretion processes envisioned here \citep{Crutcher1999,Bourke2001}.

\begin{figure}
	\includegraphics[width=\columnwidth]{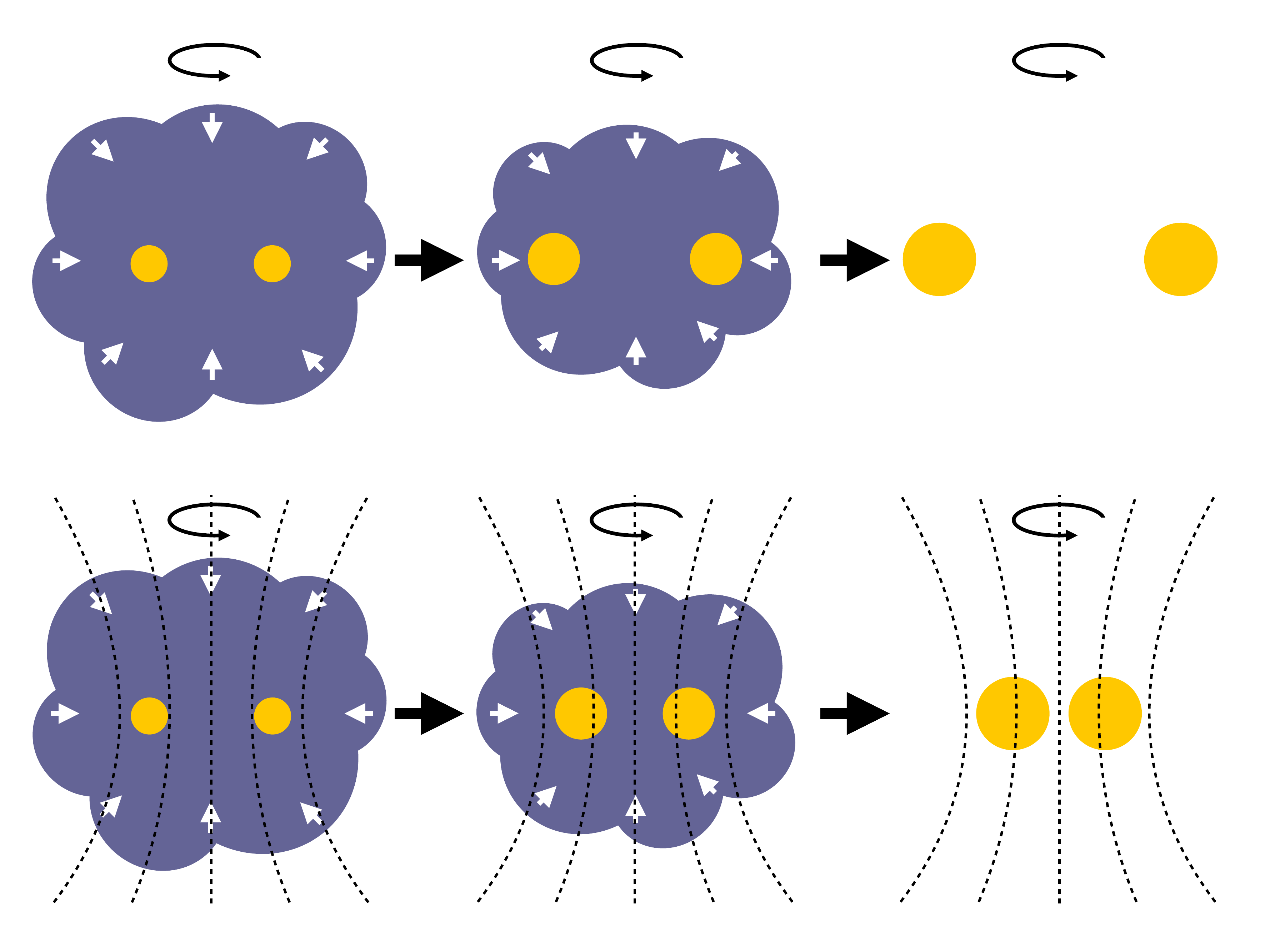}
	\caption{Accretion of material onto a wide low-mass binary with (bottom) and without (top) the presence of a magnetic field. The masses of the binary stars will increase in either case, but the final separation of the stars is more likely to decrease if a magnetic field has stripped the system of some of its angular momentum.}
	\label{fig:1}
\end{figure}

We develop a semi-analytic accretion model including the effects of magnetic braking and the removal of angular momentum from the infalling gas. Our aim is to investigate what initial conditions are needed to produce close high-mass binary systems through accretion. The main focus will be on the ability of magnetic fields  to transport angular momentum out of the molecular cloud surrounding the binary such that the binary can accrete the mass without the associated angular momentum initially contained in the gas. The accretion of low-angular momentum will drive the system to smaller separations and higher masses.  To visualise this idea see Figure \ref{fig:1}.

In section 2 we discuss alternative formation mechanisms for high-mass binaries. Section 3 describes how our accretion model is set up, and how turbulence and magnetic fields are implemented. Section 4 outlines the main results, comparing the final binary mass and separation for a range of initial parameters with and without a magnetic field present. 

\section{Formation of high-mass binaries}

A binary with both high mass and small separation can not form directly from the collapse and fragmentation of a molecular cloud since the resulting masses of the binary stars are proportional to their separation \citep{Bate1997,Bonnell1999}. We can estimate the possible masses and separations from a fragmentation process via the Jeans mass and radius, the minimum mass and radius for a fragment to be gravitationally bound considering only thermal support.
The jeans mass and radius are given by 
\begin{equation}
\label{Jeans_mass}
M_J = \left(\frac{5 R_g T}{2 G \mu}\right)^{3/2} \left(\frac{4}{3}  \pi \rho \right)^{-1/2},
\end{equation}
and 
\begin{equation}
R_J = \left(\frac{5 R_g T}{2 G \mu}\right)^{1/2} \left(\frac{4}{3}
\pi \rho \right)^{-1/2}, 
\end{equation}
where $\rho$ is the gas density, $T$ is the gas temperature, $R_g$ is the gas constant, $G$ is the gravitational constant, and
$\mu$ is the mean molecular weight. Typical values of the Jeans mass and radius are $\approx 1\, \rm{M_{\odot}}$ and $\approx 10^4$ au corresponding to densities of $10^{-19}$ g cm$^{-3}$ and temperatures of $10$ K.
To form a binary system, two fragments would have to be separated by twice their Jeans radius, such that there is a direct relationship between the mass and minimum separation of the fragments
\begin{equation}
R_{\rm sep} \propto \frac{M_*}{T},
\end{equation}
such that for reasonable gas temperatures, fragmentation can only form close binary systems with very low masses \citep{Bonnell1994}.
If fragmentation of molecular clouds were the only way to form high-mass close binary systems, then typical separations would be expected
to be in the range of $10^4$ to $10^5$ au. 

\begin{figure}
	\includegraphics[width=\columnwidth]{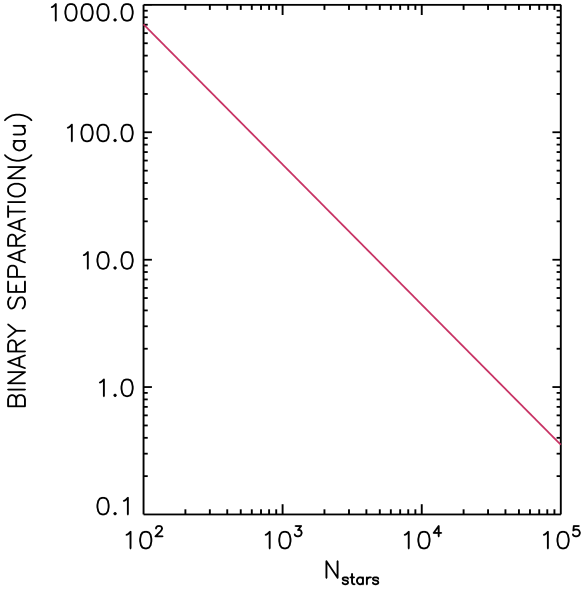}
	\caption{The minimum separation of two stars in a binary system can be estimated by equating the energies of the binary and the full cluster. This minimum separation relies on the binary absorbing all the binding energy of the cluster. Smaller separations will require alternative mechanisms than pure gravitational dynamics.}
	\label{fig:Nstars}
\end{figure}

Binary hardening in a stellar cluster is an alternative formation mechanism for high-mass close systems. Hard binaries in a cluster tend to become closer (harder) through stellar interactions \citep{Heggie1975}. This process conserves energy (in the absence of direct collisions or tidal effects) such that there is a a limit to how close a binary system can become due to stellar interactions. Ultimately, the binary can only become so close that it absorbs all of the binding energy of the cluster ($E_{binary}\sim E_{cluster}$), and in so doing forces the rest of the cluster to dissolve. The minimum separation is then given by the original binding energy of the cluster from
\begin{equation}
\frac{G M_1 M_2}{R_{\rm sep}} \approx \frac{G M_{\rm clust}^2}{ R_{\rm clust}},
\end{equation}
or 
\begin{equation}
R_{\rm sep} \ge \frac{ M_1 M_2}{M_{\rm clust}^2} R_{\rm clust}. 
\label{eq:Rsep}
\end{equation}
We can estimate the binary properties by relating the expected highest mass star to the cluster mass \citep{Larson1982,Larson2003b,Elmegreen2000,Weidner2010}. Here we will use the empirical expression derived by \cite{Larson1982,Larson2003b}:
\begin{equation}
M_{\rm star, max} \sim 1.2 M_{\rm clust}^{0.45},
\end{equation}
and assuming the most-massive star in the cluster is the primary binary star and the secondary star is 0.75 times the mass of the primary. We then use Equation \ref{eq:Rsep} to estimate the minimum separation of the binary given only the mass and size of its parent cluster. For a cluster of 1000 M$_{\odot}$ in a radius of 0.5 pc, stellar dynamics alone can form a high-mass binary system consisting of a 27 M$_{\odot}$ and a 20 M$_{\odot}$ with a minimum separation of 56 au. Figure \ref{fig:Nstars} shows the minimum binary separation given the number of stars in the cluster, $N_{\rm stars}$, for a cluster with $R_{\rm clust}=0.5$ pc and $M_{\rm clust}=N_{\rm stars}$M$_\odot$. Forming closer, or higher-mass binaries requires an alternative mechanism.

Accretion onto lower-mass systems can help form high-mass binaries \citep{Bate2000,Bonnell2005,Larson2003}, but if the accretion contains significant angular momentum as would be expected in a rotating system, then forming close systems is problematic (see below). In this paper we explore how while mass accretion  proceeds, magnetic braking can limit the accretion of angular momentum.

\section{Method}

\subsection{The initial setup}
We begin with a wide solar mass binary star system; two 1 M$_\odot$ stars separated by 100 au. In the model the binary is approximated as a point mass of 2 M$_\odot$ with an angular momentum pointing along the rotation axis, $z$. Using Kepler's 3rd law and the centre of mass condition the expression for the angular momentum of a binary can be written as:
\begin{equation}
L_{zb}=\sqrt{GM_b^3s}\frac{q}{(1+q)^2}=\frac{1}{4}\sqrt{GM_b^3s},
\label{eq:1}
\end{equation}
where $M_b$ is the total mass of the binary, $s$ is the separation, and $q$ is the mass ratio of the two stars. To simplify the system we assume a constant mass ratio of 1. In which case its separation is completely defined by the mass and angular momentum of the binary. The binary is surrounded by a spherical cloud from which material can fall in and be accreted. Any gas within an accretion radius $R_{\rm{acc}}$ equal to the binary separation, $s$, falls onto the binary. The clouds are made to either have uniform density, $\rho=\rho_0$, or to be centrally condensed, $\rho\propto R^{-2}$.   

We model the collapse as an inside out process and consider successive shells in the cloud. For each shell we sample a large number of gas elements of equal mass $m_{e}$ with random positions (x,y,z) within the shell. In total we use $10^6$ elements spread across $10^4$ shells. The cloud is given a solid body rotation around the same axis of rotation as the binary, with an angular velocity $\Omega_{\rm{cloud}}$. This means the initial velocity of any gas element is set as 
\begin{equation}
\vec{v}=\vec{\Omega}\times\vec{R}, 
\label{eq:vel}
\end{equation}
with turbulence potentially added, see section 3.2. Each element will have a different angular momentum depending on its position and its velocity given by
\begin{equation}
\vec{L}_{e}=m_{e}(\vec{R}\times\vec{v}).
\label{eq:2}
\end{equation}

We determine the fate of each element by tracking their path through first order Euler integrations of the element's position and velocity. The size of each time step, $\Delta t$, decreases as the gas element moves closer to the binary and as the total velocity increases, according to
\begin{equation}
\Delta t=|10^{-3}\frac{R}{v}|. 
\label{eq:timestep}
\end{equation}
The time steps are scaled by $10^{-3}$, a value chosen for numerical expediency. The integration stops after one of the following events occur: the gas element escapes the system, settles in a disk surrounding the binary or accretes onto the binary. The element is removed if it leaves the cloud with a velocity great enough to escape the system completely. We define a critical radius, $R_c$, where the force from the element's rotational motion,  
\begin{equation}
\vec{F}_c=\frac{m_{e}v^2_{\phi e}}{R}\hat{r},
\label{eq:3}
\end{equation}
and the strength of gravity,
\begin{equation}
\vec{F}_g=-\frac{GM_bm_{e}}{R^2}\hat{r}
\label{eq:4}
\end{equation}
are balanced. 
\begin{equation}
R_c=\frac{J^2_{ze}}{GM_b},
\label{eq:5}
\end{equation}
this is the radius where the element would settle in a stable orbit, it is only dependent on the total mass of the binary system and the z component of the specific angular momentum of the gas element, $J_{ze}$. If this critical radius is larger than the accretion radius, $R_c>R_{\rm{acc}}$, then once the element reaches $R_c$ we place it in a disk surrounding the binary stars. If however the critical radius is equal to or smaller than the accretion radius, 
\begin{equation}
R_c\leq R_{\rm{acc}},
\label{eq:6}
\end{equation}
the element is accreted onto the binary once it passes $R_{\rm{acc}}$. Equation \ref{eq:6} can be rewritten in terms of the specific angular momentum of the element and the binary as
\begin{equation}
|J_{ze}|\leq|4J_{zb}|.
\label{eq:7}
\end{equation}
Equation \ref{eq:7} shows the relative sizes of the specific angular momentum is the only factor that determines whether or not the gas element is able to accrete onto the binary. The factor of 4 comes from the assumption that the two binary stars are always equal in mass, $q=1$, in Equation \ref{eq:1}.

If accretion occurs the mass of the element is added to the binary mass,
\begin{equation}
M_{b_{\rm{new}}}=M_b+m_e,
\label{eq:8}
\end{equation}
and its angular momentum is added to the total angular momentum of the binary system,
\begin{equation}
\vec{L}_{b_{\rm{new}}}=\vec{L}_{b}+\vec{L}_{e}.
\label{eq:9}
\end{equation}
As we assume a constant mass ratio of 1, we do not differentiate between the binary mass and the mass of the two components. A new binary separation is calculated, this is also the new accretion radius,
\begin{equation}
s_{\rm{new}}=R_{\rm{acc}_{\rm{new}}}=16\frac{L_{{zb}_{\rm{new}}}^2}{GM^3_{b_{\rm{new}}}},
\label{eq:10}
\end{equation}
hence the accretion criterion, Equation \ref{eq:6}, evolves along with the binary system.

Once every gas element in a shell has escaped the system or accreted onto the binary or the surrounding disk we move on to the next shell and the process repeats. 

\subsection{Turbulence}

Turbulence is introduced to the model by randomizing the Cartesian velocity components of the gas elements according to a Gaussian distribution. The mean velocities in the x, y and z directions correspond to the overall solid body rotation of the cloud (Equation \ref{eq:vel}), and the standard deviation $\sigma$ is given by the following relation \citep{Larson1981,Heyer2004},
\begin{equation}
\sigma=0.8\,\rm{km}\,\rm{s}^{-1}\sqrt{\frac{R}{1\,\rm{pc}}}.
 \label{eq:11}
\end{equation}

\subsection{Magnetic fields}

When adding a magnetic field to the system the only impact is on ions \citep{Stahler2004}. The model estimates the overall magnetic effect by having the field act on all the material, but multiplying by the ionisation fraction \citep{Elmegreen1979,McKee1989},
\begin{equation}
I_f\approx 10^{-7}\sqrt{\frac{10^4\rm{cm}^{-3}}{n}},
\label{eq:12}
\end{equation}
$n$ is the number density of the cloud. $I_f$ effectively acts as a coupling coefficient.

We assume a uniform field threads the molecular cloud along the direction of the rotation axis, $B_0=B_z$, this is the only orientation we consider. The cloud is rotating with $v_{\phi}$, which causes the field lines to be dragged azimuthally \citep{Stahler2004}. By equating this rotational velocity, $v_{\phi}$, with the Alfven velocity,
\begin{equation}
v_A=\frac{B_0}{\sqrt{\mu_0\rho}},
\label{eq:13}
\end{equation}
acting along the field lines, the new azimuthal field component is estimated as
\begin{equation}
B_{\phi}=B_0\frac{v_{\phi}}{v_A}.
\label{eq:14}
\end{equation}
We impose an upper limit on the azimuthal field strength of $ B_{\phi}=B_0$, as at this level non-ideal processes such as diffusion and reconnection are likely to become important. We also use a constant $B_0$, assuming that non-ideal processes offset the growth of the magnetic field during infall. Both these assumptions assure a lower limit on the effects of magnetic braking.

This bending of the field lines results in a magnetic torque $\tau$ \citep{Masson2016}. The presence of a torque means there is a rate of change of angular momentum, as 
\begin{equation}
\vec{\tau}=\frac{d\vec{L}}{dt}.
\label{eq:15}
\end{equation}
The angular momentum of the cloud propagates along the magnetic field lines by Alfven waves and, in time, is removed from the system \citep{Masson2016}. For this scenario the magnetic torque per unit area is given by
\begin{equation}
T=\frac{B_zB_{\phi}R}{2\pi}.
\label{eq:16}
\end{equation}
Since the effect the torque has on each gas element depends on radial position and the time over which it acts we strip away angular momentum after each step of the integration scheme used to track the path of the gas elements. We assume it is only the z-component of the angular momentum that is being reduced by the torque: 
\begin{equation}
{L}_{ze,\rm{new}}={L}_{ze}-I_f{\tau}(R)\Delta t, 
\label{eq:integration}
\end{equation}
and we update the x- and y-velocity components accordingly before the next integration step.

We tested two field configurations; a variable $B_\phi$ described by Equation \ref{eq:14}, and a constant field, $B_\phi= B_0$. There was only a slight difference in results between the two, hence the constant field was chosen for all simulations presented in this paper. 

 \section{Results}

In our model the only factor determining whether or not gas is able to accrete (Equation \ref{eq:7}), and the resulting change in binary separation (Equation \ref{eq:10}), is its specific angular momentum. There are three regimes;  
\begin{enumerate}
  \item $|J_{ze}|>|4J_{zb}|$ : no accretion. 
  \item $|J_{ze}|\leq|4J_{zb}|$ : accretion.
  \item $|J_{ze}|\leq|1.5J_{zb}|$ : accretion and decreased separation. 
\end{enumerate}
It is possible for the two stars to gain mass, but move further apart as a consequence \citep{Bate1997}. To end up with a close and massive binary system through accretion the infalling gas must have a low specific angular momentum.

  \begin{figure*}
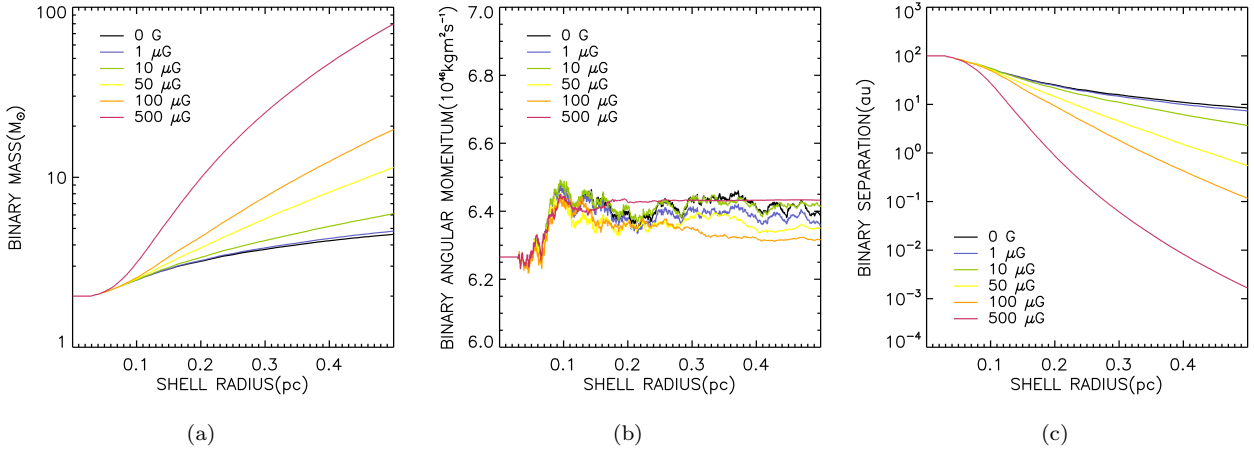
  
 \subfloat[]
	{\includegraphics[width=0.666\columnwidth]{radmasscartesian}}
     \subfloat[]
	{\includegraphics[width=0.666\columnwidth]{radangcartesian}}
    \subfloat[]
	{\includegraphics[width=0.666\columnwidth]{radsepcartesian}}  
	\caption{Change in binary mass, angular momentum and separation as consecutive shells of a uniformly dense and turbulent molecular cloud collapses. The different coloured lines correspond to placing the clouds in magnetic fields of different strengths. $M_{\rm{cloud}}=500\,\rm{M}_\odot$, $R_{\rm{cloud}}=0.5$ pc and $\Omega_{\rm{cloud}}=3\times 10^{-14}\,\rm{rad}\,\rm{s}^{-1}$.}
	\label{fig:3}
\end{figure*}

\begin{figure*}
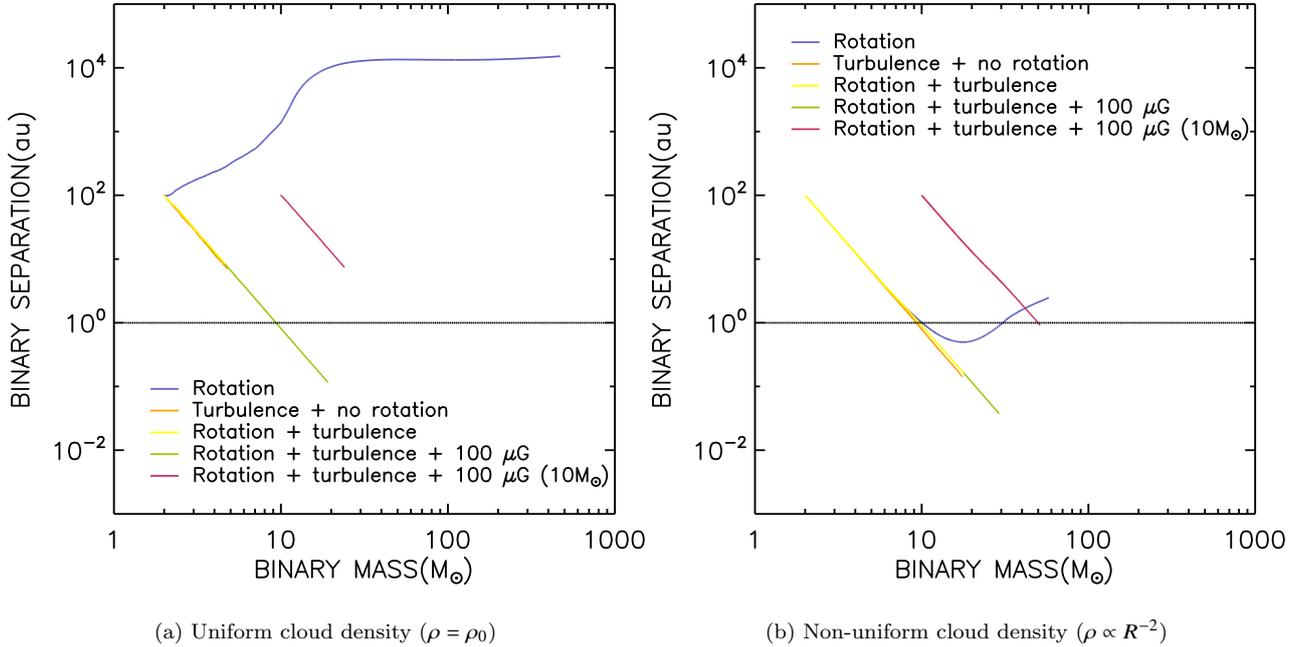

 \subfloat[Uniform cloud density ($\rho=\rho_0$)]
	{\includegraphics[width=\columnwidth]{masssepcartesian.pdf}}
     \subfloat[Non-uniform cloud density ($\rho\propto R^{-2}$)]
	{\includegraphics[width=\columnwidth]{masssepcartesiandense.pdf}}
	\caption{Change in binary mass and separation during accretion. Different coloured lines correspond to placing the binary system in different environments. For the red line the initial mass of the binary is 10 M$_\odot$, in every other scenario it is 2 M$_\odot$. The black horizontal line marks a separation of 1 au, below which binary stars are likely to merge. M$_{\rm{cloud}}=500\,\rm{M}_\odot$, $R_{\rm{cloud}}=$0.5 pc, $\Omega_{\rm{cloud}}=3\times 10^{-14}\,\rm{rad}\,\rm{s}^{-1}$ and $B_0=100\,\mu$G. }
	\label{fig:5}
\end{figure*}

In Figures \ref{fig:3} and \ref{fig:5} we place the binary stars in a molecular cloud with a mass of 500 M$_\odot$, a radius of 0.5 pc and an angular velocity of $3\times 10^{-14}\,\rm{rad}\,\rm{s}^{-1}$. These parameter values are chosen as an example of typical properties of clumps where high-mass star formation occurs \citep{Goodman1993,Urquhart2014}. 

By comparing the gravitational binding energy,
\begin{equation}
E_{g}=\frac{3GM_{\rm{cloud}}^2}{5R_{\rm{cloud}}},
\label{eq:Egrav}
\end{equation}
of the cloud with its rotational energy,
\begin{equation}
E_{\rm{rot}}=\frac{1}{2}I\Omega_{\rm{cloud}}^2,
\label{eq:Erot}
\end{equation}
we get an estimate of their relative dynamical importance in this particular scenario. In this calculation we are treating the molecular cloud as a uniformly dense full sphere, neglecting the empty central region within the initial accretion radius, corresponding to $\sim$0.1\% of the total radius. Assuming a uniform spherical distribution the moment of inertia, I, in the rotational energy equation, can be written as
\begin{equation}
I=\frac{2}{5}M_{\rm{cloud}}R_{\rm{cloud}}^2, 
\label{eq:moment}
\end{equation}
and the energy ratio simplifies to
\begin{equation}
\frac{E_{\rm{rot}}}{E_g}=\frac{R_{\rm{cloud}}^3\Omega_{\rm{cloud}}^2}{3GM_{\rm{cloud}}}.
\label{eq:ratio}
\end{equation}
For our chosen parameters $E_{\rm{rot}}/E_g\approx0.02$. Similarly we can evaluate the impact of turbulent kinetic energy using
\begin{equation}
E_{\rm{turb}}=\frac{1}{2}m_e\Sigma v_{\rm{turb}}^2. 
\label{eq:turbKE}
\end{equation}
We sum the kinetic energy corresponding to the turbulent motion, $v_{\rm{turb}}$, of every element in the cloud. $v_{\rm{turb}}$ is calculated by subtracting the mean velocity corresponding to solid body rotation from the total initial velocity. $E_{\rm{turb}}/E_g=0.06$, meaning gravity is dominating both rotation and turbulence, and we expect the cloud would collapse.   

Figure \ref{fig:3} shows how the binary mass (a), angular momentum (b) and separation (c) evolve as successive shells of the cloud collapse. The cloud collapses from the inside out, hence you can think of the shell radius along the x-axis as a proxy for time. The evolution of binary properties during accretion changes depending on the strength of the magnetic field. This is equivalent to considering different levels of magnetic coupling by varying the ionisation fraction (Equation \ref{eq:12}). The magnetic field strength determines the amount of angular momentum that is stripped from the infalling gas, which impacts the amount of material that is accreted and the separation of the final binary. 

As you can see in Figure \ref{fig:3}, for the particular system in question, the central binary stars increase in mass and decrease their separation even in the case with no magnetic field present. However, the combined mass does not even reach 5 M$_\odot$ and the binary separation is almost 10 au. In other words, this is not a close high-mass system. In contrast, a magnetic field of 100 $\mu$G or more results in a binary mass exceeding 19 M$_\odot$ with a separation of less than 0.12 au. 

It is clear from Figure \ref{fig:3} that adding a magnetic field makes it easier to form binaries with higher mass and smaller separation. The effect of the magnetic field increases with increasing field strength, and in the case of the strongest field strength, 500 $\mu$G, most of the gas accreted from the cloud is stripped entirely of angular momentum. Figure \ref{fig:3}(b) shows how nearly no angular momentum is being added to the binary during accretion of material starting at radii larger than $\sim$ 0.2 pc. 

The stronger magnetic fields provide the cloud with a greater magnetic support. To evaluate whether this support is great enough to prevent the cloud from collapsing we consider the ratio of magnetic energy\citep{Hartmann1998}, 
\begin{equation}
E_{B}=\frac{1}{3}R_{\rm{cloud}}^3B_0^2,
\label{eq:EB}
\end{equation}
to the gravitational binding energy (Equation \ref{eq:Egrav}):
\begin{equation}
\frac{E_{B}}{E_g}=\frac{5R_{\rm{cloud}}^4B_0^2}{9GM_{\rm{cloud}}^2}. 
\label{eq:ratio2}
\end{equation}
For the field strengths considered in Figure \ref{fig:3}, $B_0={1,\,10,\,50,\,100,\,500}\,\mu$G, the corresponding ratios are $E_B/E_g\approx{2\times 10^{-5},\,2\times 10^{-3},\,0.05,\,0.2,\,5}$. Gravity dominates in every case until we reach the 500 $\mu$G field. For this particular cloud the critical magnetic field strength causing the two energies to balance is $\sim$ 218 $\mu$G. For the remainder of this paper we will use a magnetic field of 100 $\mu$G. 

A different way of evaluating the magnetic field strength is calculating the Alfven Mach number, $M_A$, the ratio of the local flow velocity to the local Alfven speed. This value changes with radius, but for our parameters stays sub-Alfvenic throughout the cloud, with a maximum value of $\sim$ 0.51.  

The effect of adding different levels of physics to the accretion process is studied in Figure \ref{fig:5}. The plots show the evolution of the binary mass and separation during the cloud collapse. The black horizontal line marks a separation of 1 au, below which we consider the separation to be close.

We started with a rotating cloud, then added turbulence and a magnetic field. The effect of rotation is to increase the binary's separation once the specific angular momentum of the infalling gas is greater than that of the binary, $|J_{ze}|>|1.5J_{zb}|$, see also \cite{Bate1997}. With turbulence, there is more material present in the cloud at small specific angular momenta and hence the binary separation is able to decrease with increasing mass. As the specific angular momentum increases with radius eventually the accretion stops, limiting the growth of the binary's mass and its minimum separation. Magnetic fields act to reduce the specific angular momentum of the infalling gas, allowing accretion to continue to arbitrarily large values, while the separation can decrease to the size scale of the individual stars. Hence the full model, including a magnetic field, results in the most massive close binaries. 

We explore the differences between uniform and non-uniform density clouds. As one would expect the centrally condensed cloud ($\rho\propto R_c^{-2}$) has more mass at low specific angular momenta and hence has an initial decrease in the binary separation with added mass, see Figure \ref{fig:5}. Eventually the centrally condensed cloud affects the binary in a similar manner as does the uniform density cloud, hence we restrict the rest of our studies to the case of uniform initial density.

 \begin{figure*}
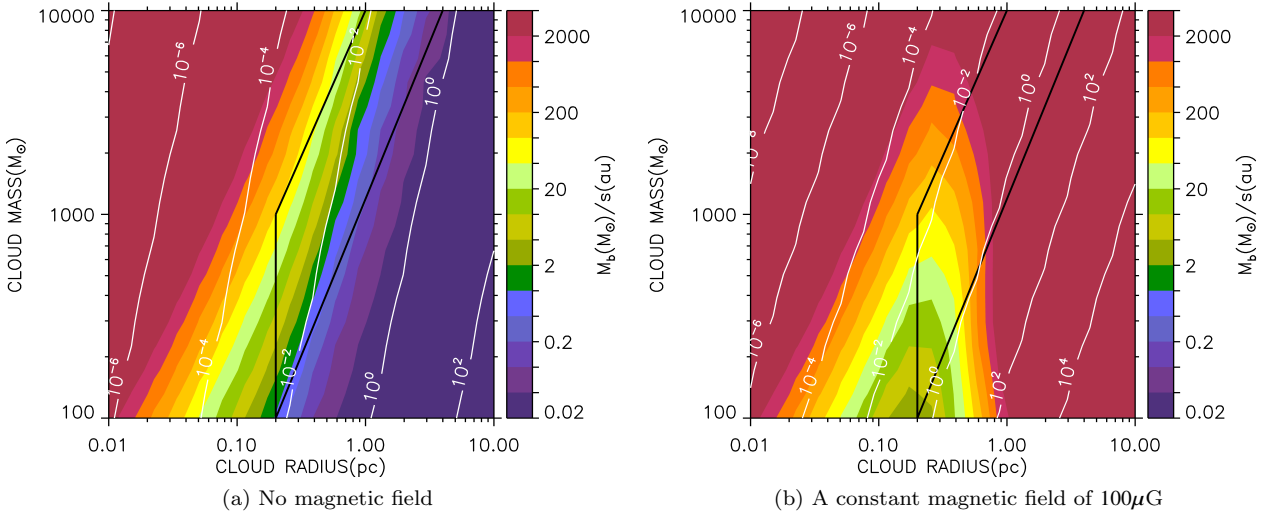

 \subfloat[No magnetic field]
	{\includegraphics[width=1\columnwidth]{mrm32cart.pdf}}
    \subfloat[A constant magnetic field of $100\mu$G]
	{\includegraphics[width=1\columnwidth]{mrm3cart.pdf}}
	\caption{The final binary mass divided by the binary separation for uniformly dense, turbulent clouds with a range of different masses and radii. The binary mass, $M_b$, is restricted to be less than 100 M$_\odot$ and the binary separation, s, to be greater than 0.01 au. We consider a close high-mass binary to be one with $M_b/s$ in excess of 20, corresponding to two 10 M$_{\odot}$ stars orbiting at 1 au. Our maximum value of $M_b/s\approx 4300$ corresponds to two 50 M$_{\odot}$ stars at a separation of approximately 5 R$_{\odot}$ (0.023 au). The black boxes outline the likely mass-radius relationship of clumps where high-mass star formation occurs \citep{Urquhart2014}. The white contour lines show $E_{\rm{rot}}/E_g$ (Equation \ref{eq:ratio}) and $E_B/E_g$ (Equation \ref{eq:ratio2}) in (a) and (b) respectively. $\Omega_{\rm{cloud}}=3\times 10^{-14}\,\rm{rad}\,\rm{s}^{-1}$.}
	\label{fig:7}
\end{figure*}

In Figure \ref{fig:7} we combine different molecular cloud parameter values and plot the resulting final mass of the binary stars divided by their separation, $M_b(\rm{M}_\odot)/s(\rm{au})$. We consider a close high-mass binary to be one with $M_b/s$ in excess of 20, corresponding to two 10 M$_\odot$ stars orbiting at 1 au. Note two 50 M$_\odot$ at a separation of 5 R$_\odot$ would give an $M_b/s\approx 4300$. For these simulations we have restricted the binary masses to be less than 100 M$_\odot$ and the binary separation to be greater than 0.01 au. 

\begin{table*}
\begin{tabular}{| c | c | c | c |c |}
Cloud radius (pc)& Binary mass (M$_\odot$) & Binary separation (au) & Magnetic field strength ($\mu$G) & Mass-to-flux ratio  \\
\hline
0.11 & 28.1 & 0.046 & 0 & $\infty$   \\
0.11 & 28.5 & 0.047 & 100 & 128.7   \\
0.25 & 14.8 & 0.297 & 0 & $\infty$   \\
0.25 & 17.8 & 0.174 & 100 & 24.9   \\
0.58 & 5.9 & 3.842 & 0 & $\infty$   \\
0.58 & 26.5 & 0.046 & 100 & 4.6   \\
1.31 & 2.8 & 35.801 & 0 & $\infty$   \\
1.31 & 43.1 & 0.010 & 100 & 0.9   \\
\end{tabular}
\caption{A comparison of resulting binary masses and separation for both magnetic and non-magnetic clouds of 1145 M$_\odot$.}
\label{tab1}
\end{table*}

We compare results when placing the binary in different molecular clouds with and without a magnetic field (100 $\mu$G), in Figure \ref{fig:7} (a) and (b) respectively. Small clouds are able to form high-mass close binaries in both the non-magnetic and magnetic cases, due to their small angular momentum. For typical cloud properties, i.e. with sizes greater than 0.2 pc, only the smallest non-magnetic clouds at a given mass form moderately high-mass close systems. For an 1145 M$_\odot$ cloud with a radius of 0.25 pc, the resultant binary has a mass of 14.8 M$_\odot$ at a separation of $\sim$ 0.3 au (see Table \ref{tab1}). Larger non-magnetic clouds form low-mass wide binaries (a 1.3 pc cloud with 1145 M$_\odot$ forms a 2.8 M$_\odot$ binary with a separation of $\sim$ 35.8 au). In contrast, while magnetic fields have little effect for small clouds, due to the smaller torques, they have significant effects on large clouds, due to the larger lever-arm and longer timescales for collapse. For example the same 1145 M$_\odot$ cloud at a radius of 1.3 pc with a 100 $\mu$G field, forms a binary system with 43 M$_\odot$ and a separation of 0.01 au. 

As noted above the size of the cloud is very important due to the resultant torque strength, which is proportional to the radius, and timescales for collapse as given by the free fall time 
\begin{equation}
t_{ff}= \sqrt{\frac{\pi^2R_{\rm{cloud}}^3}{8GM_{\rm{cloud}}}}.
\label{eq:tff}
\end{equation}
For smaller clouds, this means the time over which the torque can act is short. The black boxes in Figure \ref{fig:7} outline the approximate region of interest where observed molecular cloud clumps associated with high-mass star formation are located \citep{Urquhart2014}. Without a magnetic field this particular region is limited to the smallest clouds in its ability to form moderately high-mass or close binaries. With magnetic fields, it is increasingly easy to form very close, very high-mass binaries with increasing cloud mass or radius.

We can estimate the cloud support due to either rotation or magnetic fields by the ratio of either the rotational energy to gravitational energy or the magnetic energy to gravitational energy, respectively. The white contour lines in Figure \ref{fig:7} show these ratios indicating that most of the clouds except for those at the largest radii are dominated by gravity.    

\section{Discussion}

Our simple model shows that magnetic braking can reduce the specific angular momentum of infalling gas allowing accretion to form high-mass close binary systems. We envision this occurring in the context of a cluster formation process where fragmentation of higher mass cores produces individual stars, and the accretion of the lower density gas spread throughout the cluster environment is responsible for forming the high-mass stars located in the core of the cluster \citep{Bonnell2004,Smith2009}. In particular, \cite{Smith2009} showed that the fragmenting cores were at considerably higher gas densities, and hence more likely to be decoupled from a background magnetic field. In contrast lower density gas is likely to remain coupled to the magnetic field allowing magnetic braking to occur.

Magnetic support is commonly estimated by the mass-to-flux ratio, measured in terms of the critical value, below which the magnetic support is sufficient to prevent a gravitational collapse. The critical value of $M/\phi$ is given by:
\begin{equation}
\left(\frac{M}{\phi}\right)_{\rm crit}= \frac{c_1}{3\pi}\sqrt{\frac{5}{G}},
\label{eq:mfluxcrit}
\end{equation}
with $c_1\sim$ 0.53 determined by \cite{Mouschovias1976}. The contour lines on Figure \ref{fig:7} (b) of the energy ratio $E_B/E_g$ can be relabelled in terms of $M/\phi$, for instance the two contour lines that are approximately bounding our region of interest, $E_B/E_g$=1, 0.01, correspond to $M/\phi \approx$2, 20. In general for a magnetic field of 100 $\mu$G the mass-to-flux ratio can be written as:
\begin{equation}
\frac{M}{\phi}=0.136\left(\frac{M_{\rm cloud}}{100\, \rm M_\odot}\right)\left(\frac{R_{\rm cloud}}{1\, \rm pc}\right)^{-2},
\label{eq:mflux}
\end{equation}
in units of the critical value.

The clouds within the region of interest have mass-to-flux ratios that range from approximately 30 down to 1. From Figure \ref{fig:7}, we see that high-mass clouds are able to form close high-mass binaries even for the higher mass-to-flux ratios. For clouds of around 1000 M$_\odot$ (see Table \ref{tab1}), mass-to-flux ratios of order 5 or less result in the formation of high-mass close binaries. Observations have inferred typical mass-to-flux ratios in molecular clouds are supercritical with $M/\phi$ of approximately 2-3 \citep{Crutcher1999,Bourke2001}. Interestingly the cloud with the mass-to-flux ratio approaching the critical value, forms such a close system that it is unlikely to escape a merger.

Our simplified model neglects the presence of the low mass stars in the stellar cluster. We likewise do not consider the full density distribution and hence coupling factors that would exist in a turbulent cloud. We also neglect any increase in the magnetic field due to the collapse. Nevertheless the physical processes highlighted here should still occur and aid in the formation of close high-mass binary stars. Detailed numerical simulations of the formation of a stellar cluster including non-ideal magnetic fields, ambipolar diffusion, turbulence and stellar feedback (see below), are needed to verify our conclusions.

Feedback from star formation, in the form of ionisation, stellar winds and radiation pressure from high-mass stars, will occur simultaneously to the process described here. Simulations of feedback show that it tends to escape the cloud through weak points of low column density in the envelope \citep{Krumholz2005,Dale2005,Dale2014}. On smaller scales, accretion can occur through a disk even in the presence of feedback \citep{Kuiper2010,Kuiper2015,Kuiper2018}.

 \section{Conclusion}

The formation of high-mass close binary systems is problematic in any formation model of high-mass stars. Close systems cannot form from a direct fragmentation process due to their overlapping jeans radii. Neither can they form from a straight accretion process due to the angular momentum likely in the accreted gas. Dynamical hardening of binaries is limited by the cluster energy in which the binary is formed and is also unable to explain the origin of high-mass close binary systems. We propose that magnetic braking of the accretion flow in a cluster environment can result in the formation of high-mass close binary systems. 

In this paper we have seen that the addition of a magnetic field parallel to the direction of rotation can have a significant effect on the accretion process, allowing the accretion of mass with low specific angular momentum and hence the formation of a high-mass close binary system. We developed a toy model that considers turbulence, rotation and magnetic braking to show that typical molecular clouds with 1145 M$_\odot$, a radius of 0.58 pc and a magnetic field of 100 $\mu$G can result in the formation of a high-mass binary system with separations as close as 10 solar radii. This corresponds to a mass-to-flux ratio of 4.6 times the critical value. Hence we conclude that magnetic braking is a feasible way of forming the most massive close binary systems. 
 
\section*{Acknowledgements}
We thank Moira Jardine for helpful discussions and magnetic guidance. We would also like to thank the anonymous referee for their comments, which improved the paper. KL acknowledges financial support from the Carnegie Trust. IAB acknowledges support from the ECOGAL project, grant agreement 291227, funded by the European Research Council under ERC-2011-ADG. 




\bibliographystyle{mnras}
\bibliography{refs} 







\bsp	
\label{lastpage}
\end{document}